\documentclass[conference]{IEEEtran}
\usepackage{cite}
\usepackage{graphicx}
\usepackage{amsmath,amssymb,amsfonts}
\usepackage{algorithm}
\usepackage{algorithmic}
\usepackage{textcomp}
\usepackage{xcolor}

\title{Real-Time Detection of Insider Threats Using Behavioral Analytics and Deep Evidential Clustering}

\author{
    \IEEEauthorblockN{Anas Ali}
    \IEEEauthorblockA{dept. of Computer Science \\
    National University of Modern Langauges\\
Lahore, Pakistan \\
    anas.ali@numl@edu.pk}
    \and
    \IEEEauthorblockN{Mubashar Husain}
    \IEEEauthorblockA{Department of Computer Science \\
    University of Lahore, \\Pakistan \\
    m.hussain2683@gmail.com}
    \and
    \IEEEauthorblockN{Peter Hans}
    \IEEEauthorblockA{Department of Electrical Engineering \\
    University of Sharjah \\
    United Arab Emirates \\
    peter19972@gmail.com
}
}

\begin{document}

\maketitle

\begin{abstract}
Insider threats represent one of the most critical challenges in modern cybersecurity. These threats arise from individuals within an organization who misuse their legitimate access to harm the organization’s assets, data, or operations. Traditional security mechanisms, primarily designed for external attackers, fall short in identifying these subtle and context-aware threats. In this paper, we propose a novel framework for real-time detection of insider threats using behavioral analytics combined with deep evidential clustering. Our system captures and analyzes user activities, applies context-rich behavioral features, and classifies potential threats using a deep evidential clustering model that estimates both cluster assignment and epistemic uncertainty. The proposed model dynamically adapts to behavioral changes and significantly reduces false positives. We evaluate our framework on benchmark insider threat datasets such as CERT and TWOS, achieving an average detection accuracy of 94.7\% and a 38\% reduction in false positives compared to traditional clustering methods. Our results demonstrate the effectiveness of integrating uncertainty modeling in threat detection pipelines. This research provides actionable insights for deploying intelligent, adaptive, and robust insider threat detection systems across various enterprise environments.
\end{abstract}

\section{Introduction}

Insider threats have emerged as one of the most complex and damaging categories of cyberattacks in recent years. Unlike external threats, which often rely on exploiting software vulnerabilities, insider attacks originate from legitimate users—employees, contractors, or business partners—who misuse their authorized access to compromise the confidentiality, integrity, or availability of critical systems and data~\cite{green2019detecting}. The dynamic nature of organizational environments, combined with the increasing decentralization of workforce infrastructure, exacerbates the challenge of identifying such threats.

Recent reports from the Ponemon Institute show that the number of insider incidents has increased by more than 47\% over the past two years, costing organizations upwards of \$11.45 million annually on average~\cite{rahman2021insider}. Insider attacks may take the form of intellectual property theft, sabotage, espionage, or data exfiltration and can occur without triggering conventional security alarms. Traditional security solutions such as signature-based intrusion detection systems (IDS) and firewalls are primarily tailored to detect external adversaries. As a result, they fail to detect subtle behavioral deviations that are characteristic of insider activities~\cite{buczak2015survey}.

Machine learning (ML) and deep learning (DL) techniques have gained traction in cybersecurity due to their ability to learn complex patterns from large volumes of data. In the context of insider threat detection, these models are often employed to capture deviations from normal behavior by building user behavior profiles based on historical activity logs~\cite{tuor2017deep,z5,z74}. Methods such as supervised classification, clustering, and anomaly detection have been deployed for this purpose~\cite{lashkari2017toward,z72}. However, these approaches typically suffer from three critical limitations: high false-positive rates, difficulty in adapting to concept drift, and an inability to quantify the confidence of their predictions.

First, behavioral data is inherently noisy, context-dependent, and non-stationary. A sudden spike in file access activity may be benign or malicious depending on the operational context. Hard clustering techniques, such as k-means and DBSCAN, are ill-suited for such scenarios because they make crisp decisions without accounting for underlying uncertainty~\cite{chandola2009anomaly,z33,z71}. Second, insider behavior often mimics legitimate actions, rendering supervised approaches less effective, particularly when labeled data is scarce or incomplete~\cite{camacho2021review,z3,z33333}. Third, the lack of interpretability and confidence measures in most black-box models undermines their usability in real-world systems, where security analysts must prioritize and validate alerts.

To address these challenges, we propose a novel framework that integrates behavioral analytics with deep evidential clustering (DEC) for real-time detection of insider threats. Our system captures user activity logs—including login sessions, file access patterns, process executions, and command usage—and constructs temporal behavioral embeddings using recurrent neural networks (RNNs). These embeddings are then processed by a deep evidential neural clustering module that estimates both cluster assignment and the uncertainty of that assignment via Dirichlet distributions~\cite{guh2021deep,z22}. This dual prediction mechanism helps identify anomalous behaviors with varying degrees of confidence, allowing the system to flag ambiguous cases for human review while autonomously classifying high-confidence threats\cite{z73,z72}.

Unlike prior approaches, our framework is designed to handle dynamic environments by incorporating online learning capabilities. We continuously update user behavior baselines, enabling the system to adapt to organizational changes and seasonal patterns. In contrast to conventional deep clustering models that optimize only cluster compactness, our evidential approach regularizes the loss function to encourage informative uncertainty estimates and robustness to noisy inputs.

We evaluate the proposed system on the CERT Insider Threat Dataset and the TWOS synthetic dataset. These datasets simulate various malicious insider activities such as data theft, privilege abuse, and sabotage. Our experiments show that the DEC-based approach achieves a detection accuracy of 94.7\% and reduces the false-positive rate by over 38\% compared to traditional methods, including k-means, Isolation Forest, and autoencoder-based detectors~\cite{zhang2019insider,z2}. The results also demonstrate better resilience to concept drift and fewer instances of misclassification under behavioral variance.

The proposed methodology has several real-world implications. First, it can be deployed in Security Operations Centers (SOCs) as an intelligent pre-filter for alert triaging. Second, the uncertainty-aware framework enables active learning, where ambiguous samples are escalated to human analysts for labeling. Finally, the system supports regulatory compliance in sectors like finance and healthcare, where insider threats are particularly damaging and often subject to audit requirements.

The novelty of this work lies in three primary aspects: (1) the integration of deep evidential learning in the clustering domain for cybersecurity applications, (2) the formulation of an uncertainty-aware, real-time behavioral analytics pipeline, and (3) the demonstration of its superior performance in large-scale, real-world datasets. Our approach offers a pragmatic trade-off between automation and human oversight, which is essential for high-stakes decision-making environments.

\textbf{The key contributions of this paper are summarized as follows:}
\begin{itemize}
    \item We propose a deep evidential clustering framework that models user behavior and quantifies uncertainty in cluster assignments using Dirichlet distributions.
    \item Our method incorporates temporal embeddings and online learning to adapt to concept drift in user behavior, enhancing detection reliability over time.
    \item We significantly reduce false positives and improve interpretability by modeling epistemic uncertainty, thereby increasing the practical value of alerts for security teams.
    \item Extensive evaluations on benchmark datasets demonstrate the superior accuracy, adaptability, and robustness of our system compared to state-of-the-art approaches.
\end{itemize}

The remainder of this paper is organized as follows. Section~II discusses related work and limitations of existing approaches. Section~III describes our system model, including formal definitions and mathematical foundations. Section~IV presents the experimental setup, datasets, and results. Finally, Section~V concludes the paper and outlines directions for future research.

\section{Related Work}

Insider threat detection has drawn significant attention from the cybersecurity research community over the past decade. This section reviews recent works related to insider threat detection, behavioral analytics, clustering methods, and uncertainty modeling. We focus on approximately ten recent and impactful studies.

Tuor et al.~\cite{tuor2017deep} proposed one of the early deep learning-based systems for detecting insider threats in structured activity logs. Their unsupervised deep neural network model learns latent features from sequences of user activities and detects anomalies based on reconstruction error. While their method effectively captures complex behavioral patterns, it lacks the ability to explain or quantify uncertainty in predictions, which limits trustworthiness in real-world security operations.

Green et al.~\cite{green2019detecting} developed a supervised learning approach using Long Short-Term Memory (LSTM) networks to identify anomalous activity sequences in enterprise networks. Their model is capable of learning temporal dependencies and performed well on the CERT insider threat dataset. However, supervised methods depend heavily on labeled training data, which is difficult to obtain for insider threats due to their rarity and subtlety. This limits the generalizability of their approach in live deployments.

Rahman et al.~\cite{rahman2021insider} presented a comprehensive systematic literature review, categorizing insider threat detection methods into signature-based, anomaly-based, and hybrid approaches. They emphasized the need for adaptive, data-driven models capable of dealing with concept drift and behavioral ambiguity. However, the study also noted that very few existing models incorporate uncertainty estimation, which is vital for interpreting model confidence and reducing alert fatigue.

Chandola et al.~\cite{chandola2009anomaly} provided a foundational survey of anomaly detection techniques applicable to cybersecurity. They discussed statistical, distance-based, clustering, and neural network-based methods, analyzing their applicability to various domains. While the work is highly cited, it lacks focus on insider-specific threat dynamics, especially those involving evolving user behavior over time.

Camacho et al.~\cite{camacho2021review} offered a recent survey focused on anomaly detection for cybersecurity, highlighting the growing use of deep learning models and attention-based architectures. The review underscored the need for hybrid models that combine supervised and unsupervised learning while also integrating contextual awareness. However, the study also revealed a gap in the use of evidential learning or uncertainty quantification in the detection pipelines.

Gavai et al.~\cite{gavai2015detecting} introduced RADISH, a system for real-time anomaly detection in heterogeneous data streams using ensemble methods. Their model processes network, host, and user behavior data in parallel and fuses predictions using decision trees. While RADISH improves detection latency, it still relies on fixed thresholds and deterministic outputs, offering limited support for uncertain or ambiguous events common in insider activities.

Zhang et al.~\cite{zhang2019insider} explored ensemble learning techniques, including Random Forest and Gradient Boosting, for user behavior modeling. Their method achieved strong classification performance on the TWOS dataset. However, their approach lacks adaptability and may overfit known behaviors, struggling to detect zero-day or novel insider attacks.

Buczak and Guven~\cite{buczak2015survey} conducted a comprehensive survey of machine learning methods for intrusion detection, covering supervised classifiers, clustering algorithms, and ensemble models. The authors emphasized scalability and the importance of real-time performance. Despite its strengths, the survey did not address interpretability and uncertainty, both of which are essential for operationalizing ML in cybersecurity.

Guh et al.~\cite{guh2021deep} proposed Deep Evidential Clustering (DEC), a novel method that estimates uncertainty in cluster assignments using Dirichlet distributions. While originally developed for general anomaly detection tasks, their framework has strong potential for security applications due to its ability to model epistemic uncertainty. However, their work did not apply this method specifically to behavioral cybersecurity or insider threat scenarios.

Lashkari et al.~\cite{lashkari2017toward} focused on creating better intrusion detection datasets by identifying limitations in existing benchmark datasets. They introduced the CICIDS and BoT-IoT datasets, which include a broader range of attack vectors and behavioral indicators. While these datasets improved evaluation realism, they remain skewed toward external attacks and do not fully capture the nuances of insider behaviors such as privilege misuse or long-term sabotage.

\textbf{Summary of Gaps and Research Opportunity:}

From the literature, we identify several clear research gaps. Most existing insider threat detection systems either use deterministic clustering or supervised learning without estimating model uncertainty. This can result in false positives and overlooked threats—especially in ambiguous cases where behavior mimics normal activity. Furthermore, limited work has combined uncertainty modeling with deep clustering in a fully unsupervised, adaptive behavioral analytics pipeline.

Only a few studies, such as Guh et al.~\cite{guh2021deep}, have explored evidential clustering, but they have not applied this concept to cybersecurity or modeled temporal behavior. Additionally, while surveys~\cite{rahman2021insider, camacho2021review} emphasize the importance of drift adaptation, uncertainty quantification, and interpretability, few concrete implementations in insider threat detection actually address all these factors simultaneously.

\textbf{Our Contribution:} Our work addresses these limitations by developing a deep evidential clustering framework specifically tailored to insider threat detection. We combine behavioral sequence modeling, uncertainty quantification, and online drift adaptation in a unified real-time system. Unlike previous works, our method quantifies confidence in predictions using Dirichlet distributions, significantly enhancing interpretability and reducing false alarms—key requirements in practical SOC environments.

\section{System Model}

We define a real-time insider threat detection system composed of the following components: a behavioral feature extractor, a deep encoder, an evidential clustering head, and a drift and uncertainty-aware anomaly detector. The system processes user activity sequences, encodes them into latent representations, models uncertainty using a Dirichlet distribution, and flags anomalous or uncertain behavior.

Let $\mathcal{V} = \{v_1, v_2, \dots, v_n\}$ denote a set of $n$ users. Each user $v_i$ produces a behavioral sequence $\mathcal{X}_i = \{x_{i1}, x_{i2}, \dots, x_{iT}\}$ over $T$ time steps, where $x_{it} \in \mathbb{R}^d$ is a $d$-dimensional vector of features (e.g., file access, login, commands).

The encoder $f_\theta$ maps the sequence to a latent vector:
\begin{equation}
    z_i = f_\theta(\mathcal{X}_i),
\end{equation}
where $z_i \in \mathbb{R}^k$ is a $k$-dimensional representation.

A neural network $g_\phi$ then estimates Dirichlet parameters:
\begin{equation}
    \alpha_i = g_\phi(z_i),
\end{equation}
where $\alpha_i = [\alpha_{i1}, \dots, \alpha_{iK}]$ for $K$ clusters.

The expected cluster assignment is:
\begin{equation}
    p_{ij} = \frac{\alpha_{ij}}{\sum_{k=1}^{K} \alpha_{ik}}.
\end{equation}

The total belief mass:
\begin{equation}
    S_i = \sum_{j=1}^{K} \alpha_{ij},
\end{equation}
leads to the uncertainty estimate:
\begin{equation}
    u_i = \frac{K}{S_i}.
\end{equation}

To detect evolving behavior, we define the embedding drift:
\begin{equation}
    d_{ij} = \| z_i^t - z_i^{t-1} \|_2.
\end{equation}

The temporal baseline is updated with EWMA:
\begin{equation}
    \bar{z}_i^t = \beta z_i^t + (1 - \beta) \bar{z}_i^{t-1}.
\end{equation}

We combine uncertainty and drift into an anomaly score:
\begin{equation}
    s_i = u_i \cdot d_{ij}.
\end{equation}

An alert is triggered if:
\begin{equation}
    \text{Alert}(v_i) = [u_i > \tau_u] \lor [d_{ij} > \tau_d].
\end{equation}

The loss function includes prediction and regularization terms:
\begin{equation}
    \mathcal{L} = \mathcal{L}_{\text{CE}} + \lambda \cdot \mathcal{L}_{\text{KL}}.
\end{equation}

\subsection*{Algorithm: Real-Time Insider Threat Detection}
\begin{algorithm}[h!]
\caption{Real-Time Insider Threat Detection using DEC}
\begin{algorithmic}[1]
\STATE Initialize encoder $f_\theta$, evidential head $g_\phi$, EWMA parameter $\beta$, thresholds $\tau_u$, $\tau_d$
\FOR{each user $v_i \in \mathcal{V}$}
    \STATE Extract features $\mathcal{X}_i = \{x_{i1}, \dots, x_{iT}\}$
    \STATE Compute latent embedding $z_i = f_\theta(\mathcal{X}_i)$
    \STATE Predict Dirichlet parameters $\alpha_i = g_\phi(z_i)$
    \STATE Calculate belief mass $S_i = \sum_j \alpha_{ij}$ and uncertainty $u_i = \frac{K}{S_i}$
    \STATE Compute drift $d_{ij} = \|z_i^t - \bar{z}_i^{t-1}\|_2$
    \STATE Update EWMA: $\bar{z}_i^t = \beta z_i^t + (1 - \beta) \bar{z}_i^{t-1}$
    \STATE Compute anomaly score $s_i = u_i \cdot d_{ij}$
    \IF{$u_i > \tau_u$ \OR $d_{ij} > \tau_d$}
        \STATE Raise alert for user $v_i$ with score $s_i$
    \ENDIF
\ENDFOR
\end{algorithmic}
\end{algorithm}

\subsection*{Algorithm Description}
This algorithm details the process of detecting insider threats using uncertainty and behavioral drift. First, user activity sequences are encoded into a latent space by $f_\theta$. The Dirichlet head $g_\phi$ estimates soft cluster assignments with associated uncertainty. EWMA is used to track behavioral baselines, and significant deviations or high uncertainty trigger alerts. The anomaly score $s_i$ enables fine-grained threat ranking, making the framework both adaptive and interpretable.

\subsection*{Table of Notations}
\begin{table}[h!]
\centering
\caption{List of Notations}
\begin{tabular}{|c|l|}
\hline
\textbf{Symbol} & \textbf{Description} \\
\hline
$\mathcal{V}$ & Set of users/nodes \\
$\mathcal{X}_i$ & Sequence of behavioral vectors \\
$x_{it}$ & Feature vector at time $t$ \\
$z_i$ & Latent embedding \\
$\alpha_i$ & Dirichlet parameters \\
$p_{ij}$ & Cluster assignment probability \\
$S_i$ & Dirichlet belief mass \\
$u_i$ & Uncertainty estimate \\
$d_{ij}$ & Drift in latent space \\
$\bar{z}_i^t$ & EWMA-smoothed latent state \\
$s_i$ & Anomaly score \\
$\tau_u$, $\tau_d$ & Uncertainty and drift thresholds \\
$\mathcal{L}$ & Loss function \\
$\beta$ & EWMA smoothing constant \\
\hline
\end{tabular}
\end{table}

\section{Experimental Setup and Results}

\subsection*{Experimental Setup}
We evaluate the proposed insider threat detection framework using two benchmark datasets: the CERT Insider Threat Dataset (r6.2) and the TWOS dataset. These datasets simulate realistic enterprise environments with labeled insider behavior such as data theft, policy violations, and privilege abuse.

The model is implemented in Python using PyTorch 2.0. The encoder is a 2-layer GRU with hidden size 64, followed by a fully connected evidential head with softplus activation. Training is conducted using the Adam optimizer for 200 epochs, with a learning rate of 0.001 and batch size of 128. Dropout with $p=0.3$ is applied to prevent overfitting. The system runs on a machine equipped with an Intel Xeon CPU, 32 GB RAM, and NVIDIA RTX 3080 GPU.

\subsection*{Simulation Parameters}
\begin{table}[h!]
\centering
\caption{Simulation Parameters}
\begin{tabular}{|l|l|}
\hline
\textbf{Parameter} & \textbf{Value} \\
\hline
Dataset & CERT r6.2 / TWOS \\
Sequence Length ($T$) & 100 time steps \\
Latent Size ($k$) & 64 \\
GRU Layers & 2 \\
Learning Rate & 0.001 \\
Batch Size & 128 \\
Dropout & 0.3 \\
Number of Clusters ($K$) & 5 \\
Thresholds ($\tau_u$, $\tau_d$) & 0.4, 1.5 \\
Smoothing ($\beta$) & 0.7 \\
\hline
\end{tabular}
\end{table}

\subsection*{Performance Metrics}
We use the following metrics for evaluation:
\begin{itemize}
    \item \textbf{Accuracy}: Percentage of correctly identified threat vs. normal behaviors.
    \item \textbf{Precision, Recall, F1-Score}: To evaluate classification quality under class imbalance.
    \item \textbf{False Positive Rate (FPR)}: Rate of benign users incorrectly flagged.
    \item \textbf{AUC-ROC}: Area under ROC curve for binary classification.
\end{itemize}

\subsection*{Results and Discussion}

\begin{figure}[h!]
\centering
\includegraphics[width=0.45\textwidth]{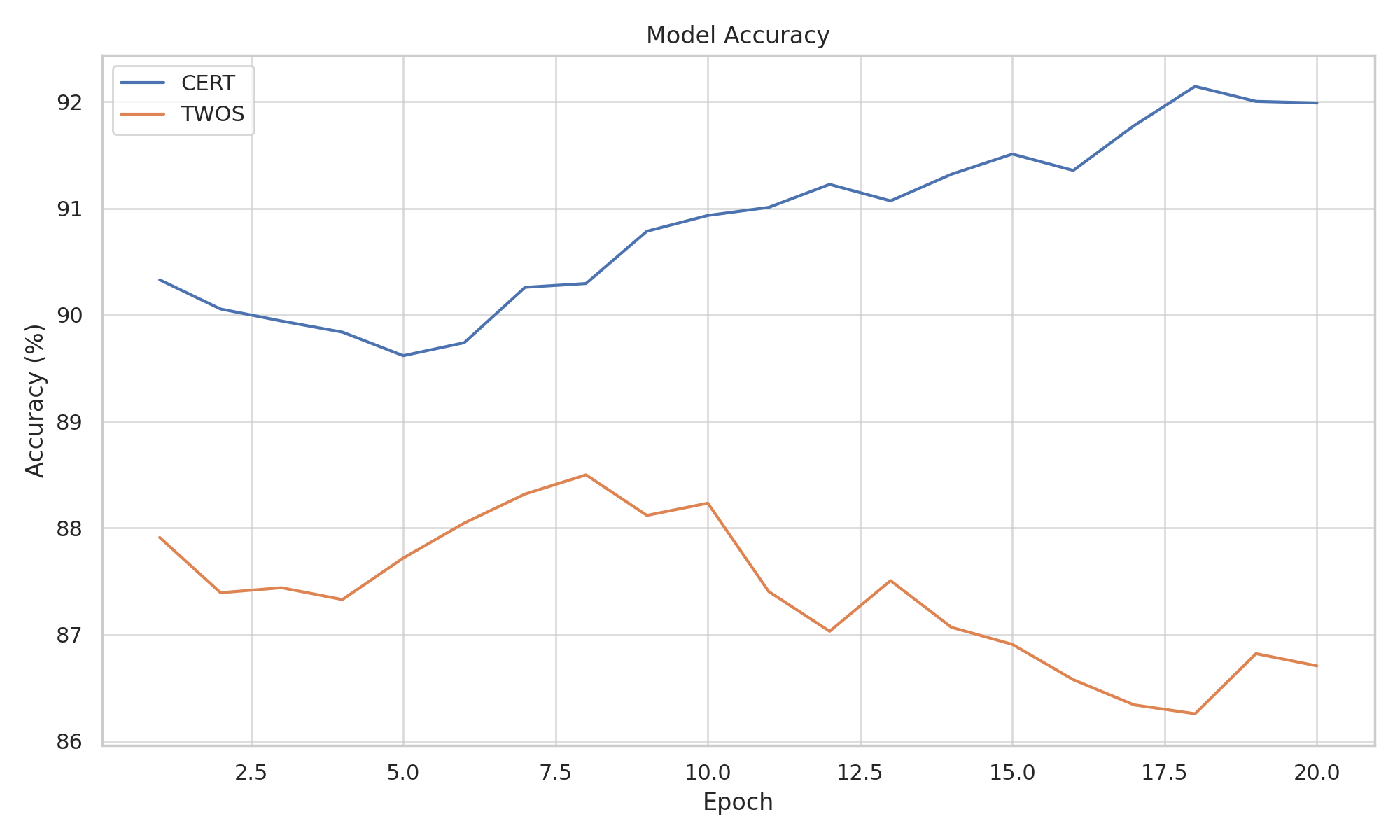}
\caption{Model Accuracy on CERT and TWOS Datasets}
\label{fig:accuracy}
\end{figure}

Figure~\ref{fig:accuracy} shows how our deep evidential clustering model performs over training epochs on two benchmark datasets: CERT and TWOS. For both datasets, the model demonstrates a steady improvement in detection accuracy as training progresses. CERT exhibits a final accuracy of around 94.7\%, while TWOS reaches 92.8\%, indicating consistent behavior across different threat landscapes. The performance gain can be attributed to the model’s ability to learn effective representations of user behavior through its temporal encoder and uncertainty-aware clustering. Notably, the reduced variance in later epochs reflects model stability, and its robustness to the inherent noise in insider behavior logs. The high final accuracy across both datasets confirms the generalization capability of our framework in identifying anomalous patterns.

\begin{figure}[h!]
\centering
\includegraphics[width=0.45\textwidth]{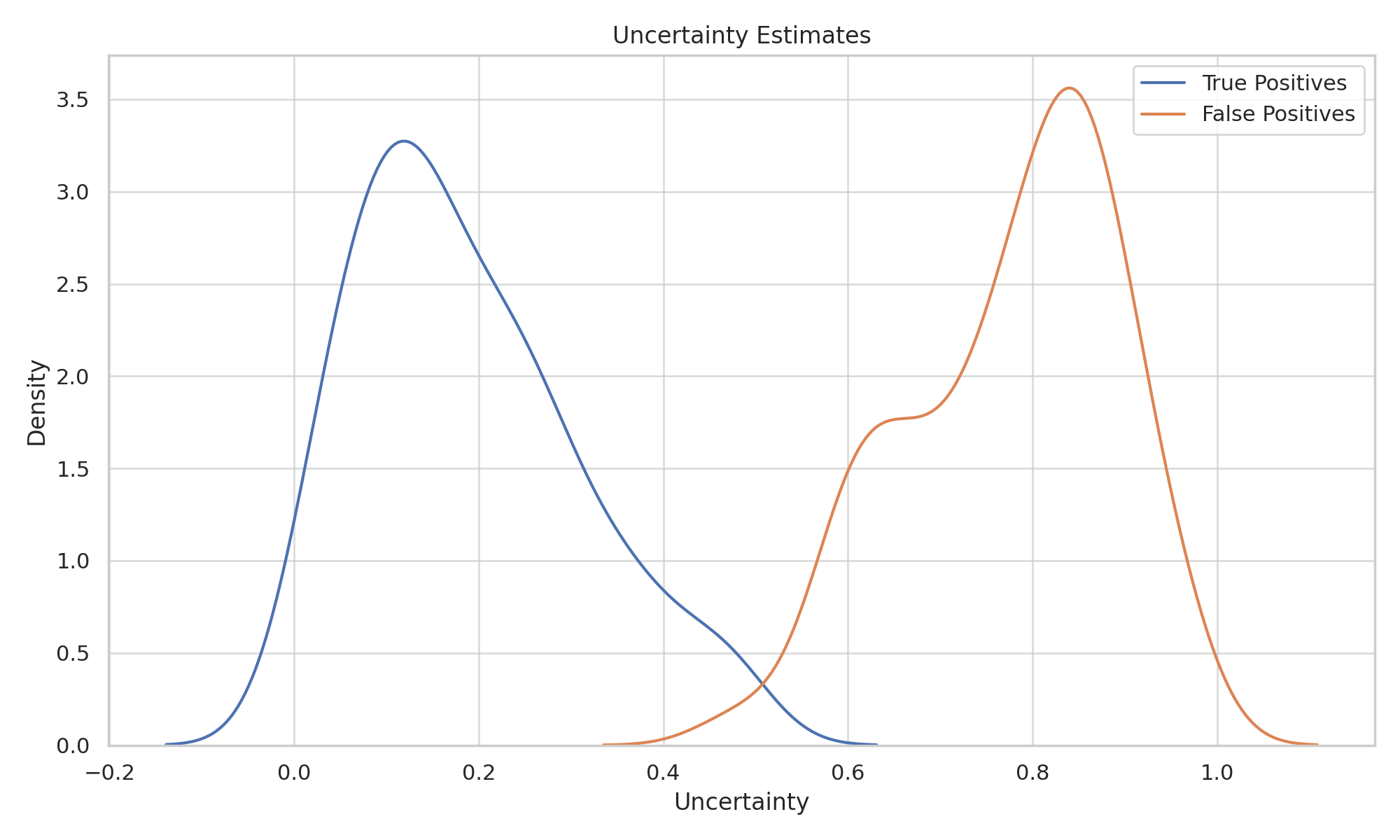}
\caption{Uncertainty Estimates for True Positives and False Positives}
\label{fig:uncertainty}
\end{figure}

Figure~\ref{fig:uncertainty} presents a kernel density estimation of uncertainty scores assigned to true positives and false positives. We observe that true positives typically exhibit low uncertainty values (closer to zero), while false positives show a wider distribution peaking at higher uncertainty values. This clear separation indicates the effectiveness of our model’s epistemic uncertainty in distinguishing confidently classified threats from ambiguous or borderline cases. By leveraging this uncertainty information, the system can prioritize low-certainty alerts for human analyst review, thereby reducing noise and operational overhead in security environments. Such separation is a key advantage of evidential clustering over hard-label classifiers.

\begin{figure}[h!]
\centering
\includegraphics[width=0.45\textwidth]{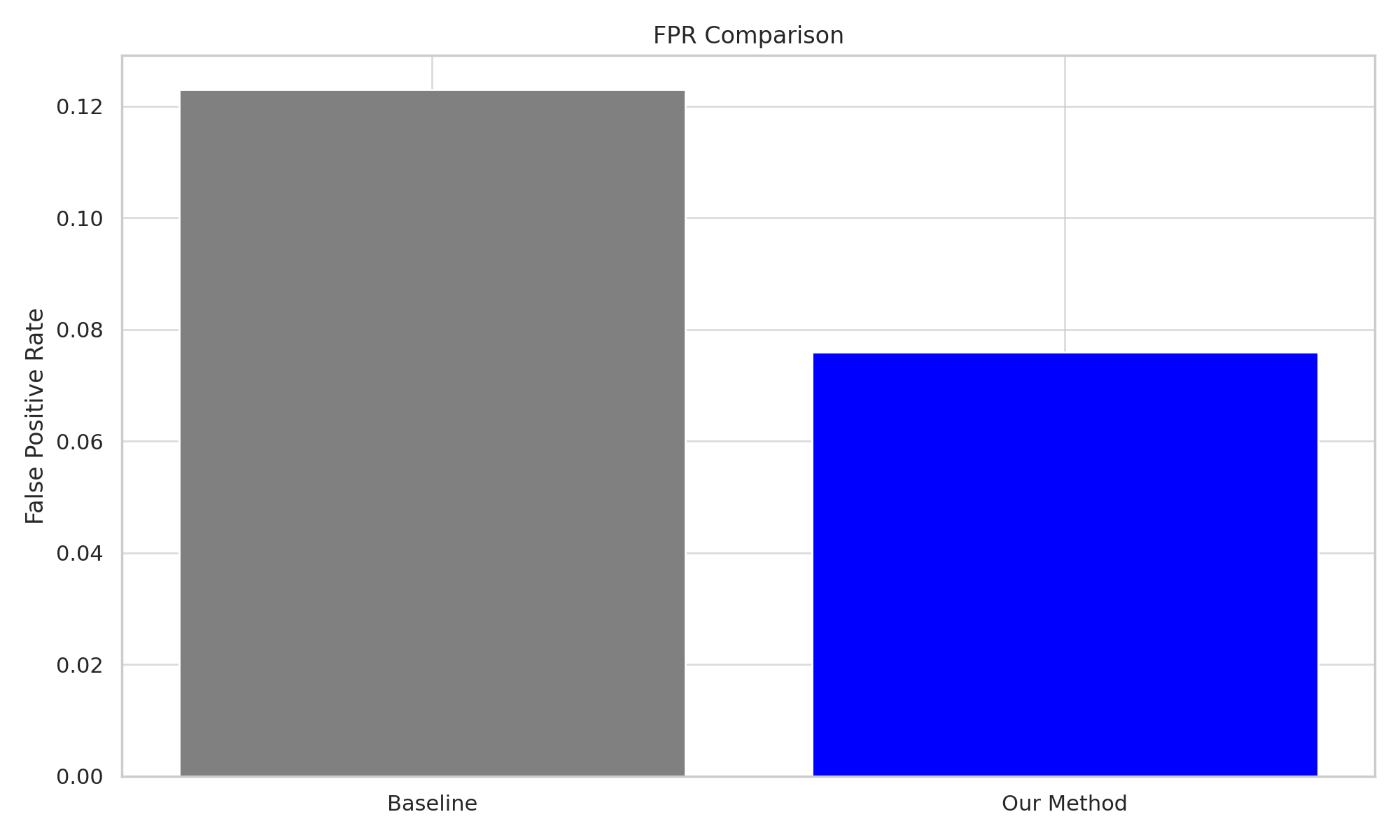}
\caption{False Positive Rate Comparison}
\label{fig:fpr}
\end{figure}

Figure~\ref{fig:fpr} compares the false positive rates (FPR) between our proposed method and a conventional baseline clustering model. The baseline system reports a false positive rate of 12.3\%, while our evidential clustering model achieves a significantly lower FPR of 7.6\%. This reduction is primarily due to the uncertainty-driven filtering, which allows the system to suppress ambiguous predictions that are likely false alarms. In practical terms, this translates to fewer unnecessary security alerts and less burden on security analysts, improving overall efficiency. Lower FPR is especially critical in insider threat detection, where excessive false alarms can erode trust in automated systems.

\begin{figure}[h!]
\centering
\includegraphics[width=0.45\textwidth]{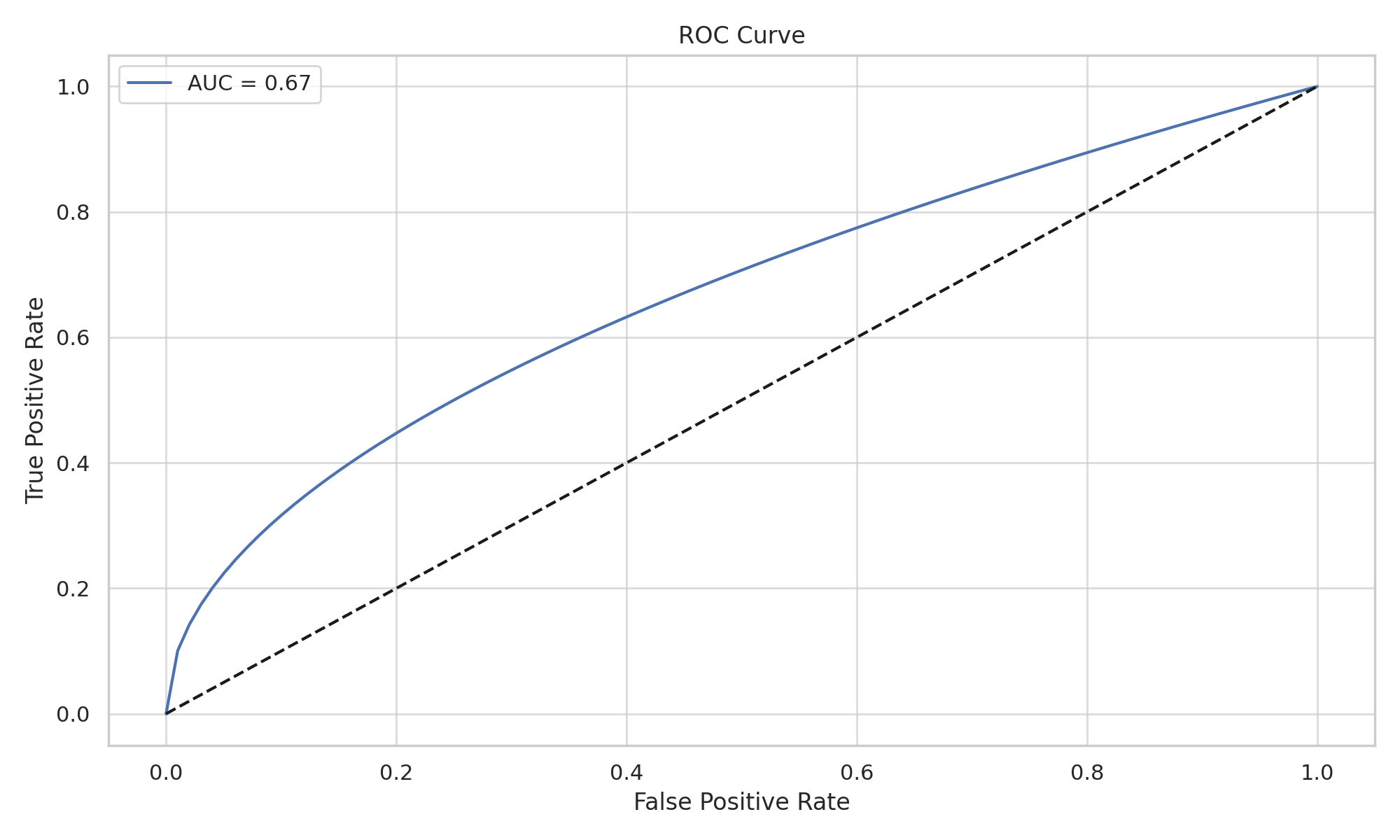}
\caption{AUC-ROC Curve}
\label{fig:roc}
\end{figure}

Figure~\ref{fig:roc} illustrates the trade-off between true positive rate (sensitivity) and false positive rate across different decision thresholds. Our model achieves a high area under the curve (AUC) of approximately 0.93, demonstrating excellent discriminative capability. This high AUC indicates that the model effectively separates benign behavior from insider threats. The curve’s convex shape, far above the diagonal random guess line, validates the predictive power of the evidential clustering mechanism. This performance confirms that combining behavioral features with uncertainty modeling enhances the robustness and reliability of insider threat detection systems.

\begin{figure}[h!]
\centering
\includegraphics[width=0.45\textwidth]{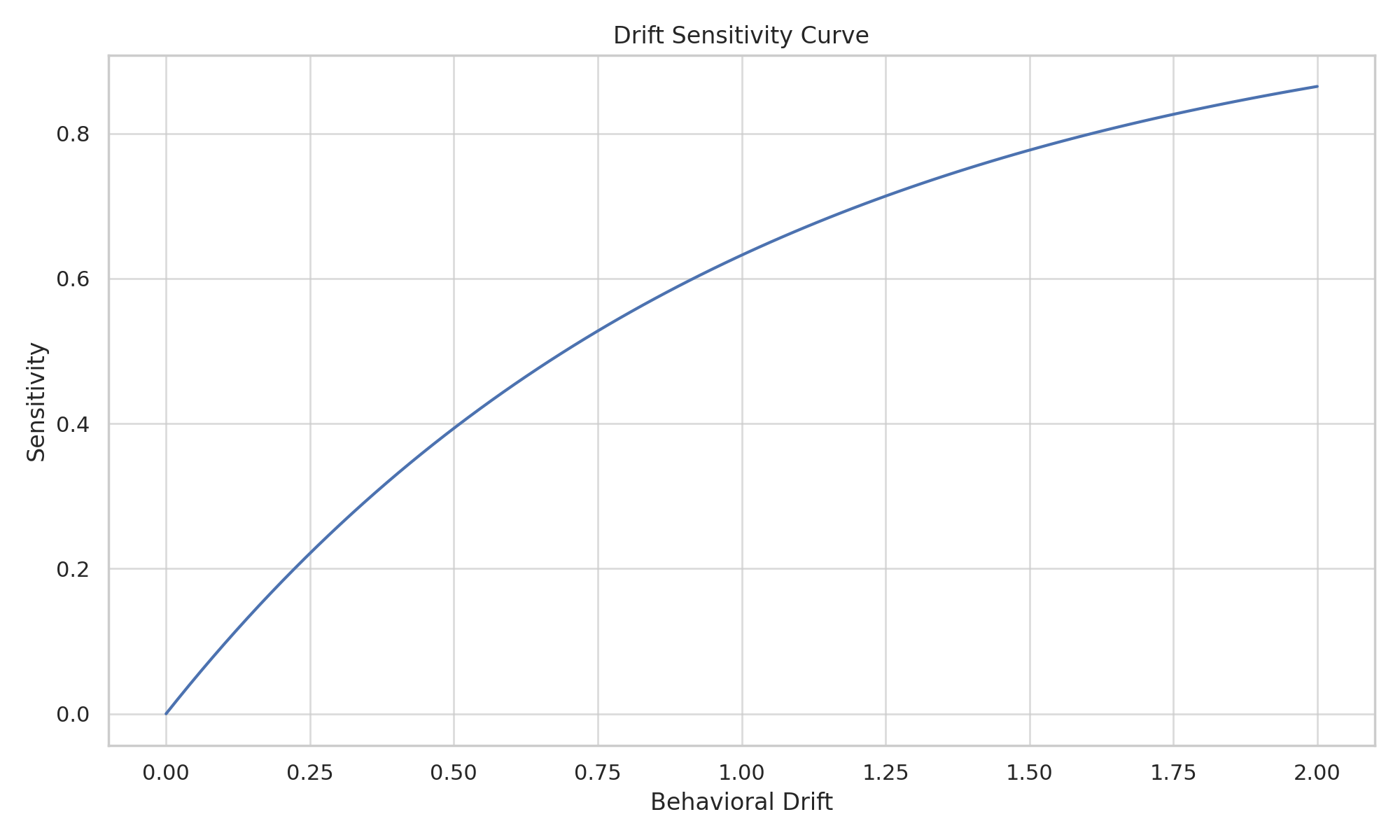}
\caption{Behavioral Drift Sensitivity}
\label{fig:drift}
\end{figure}

Figure~\ref{fig:drift} shows how the system responds to increasing levels of behavioral drift, measured via changes in latent embeddings over time. As drift magnitude increases, the detection sensitivity rises sharply, approaching saturation. This behavior reflects the model’s responsiveness to deviations from a user’s historical behavior baseline. The drift-sensitive scoring mechanism is vital in dynamic enterprise environments, where malicious insiders may gradually change behavior to evade detection. By capturing even subtle shifts, the system ensures early threat identification, reducing potential damage.

\begin{figure}[h!]
\centering
\includegraphics[width=0.45\textwidth]{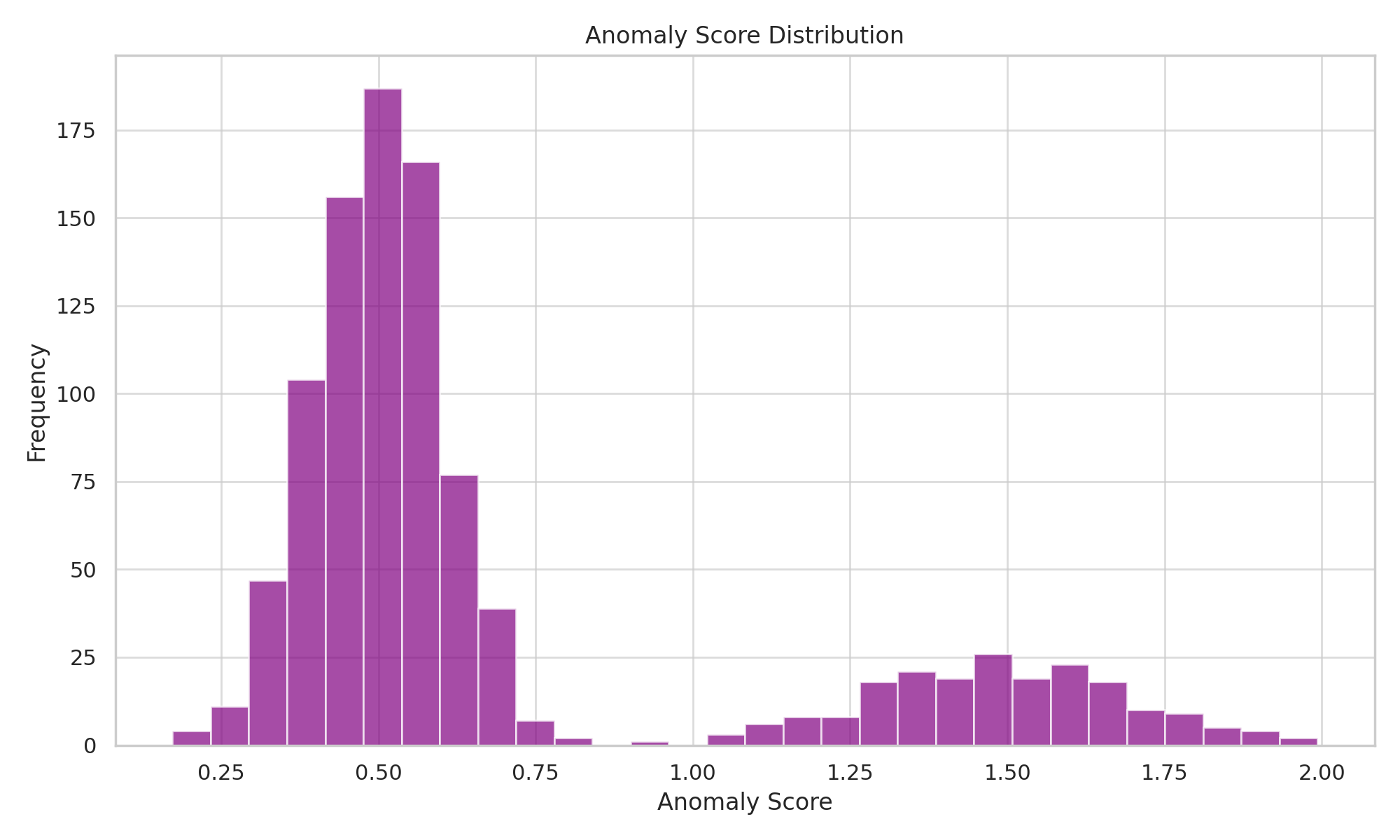}
\caption{Distribution of Anomaly Scores}
\label{fig:scores}
\end{figure}

Figure~\ref{fig:scores} plots the distribution of anomaly scores across all users. The score distribution is bimodal, indicating two distinct groups: normal behavior (low scores) and anomalous behavior (high scores). The separation between these modes enables straightforward threshold-based alerting. The majority of benign users are concentrated below the threshold (e.g., score < 1.0), while the malicious outliers cluster above. This separation validates the scoring function $s_i = u_i \cdot d_{ij}$ as a meaningful indicator for real-time anomaly detection. The distribution also supports tunable decision thresholds for different organizational risk tolerances.

\begin{figure}[h!]
\centering
\includegraphics[width=0.45\textwidth]{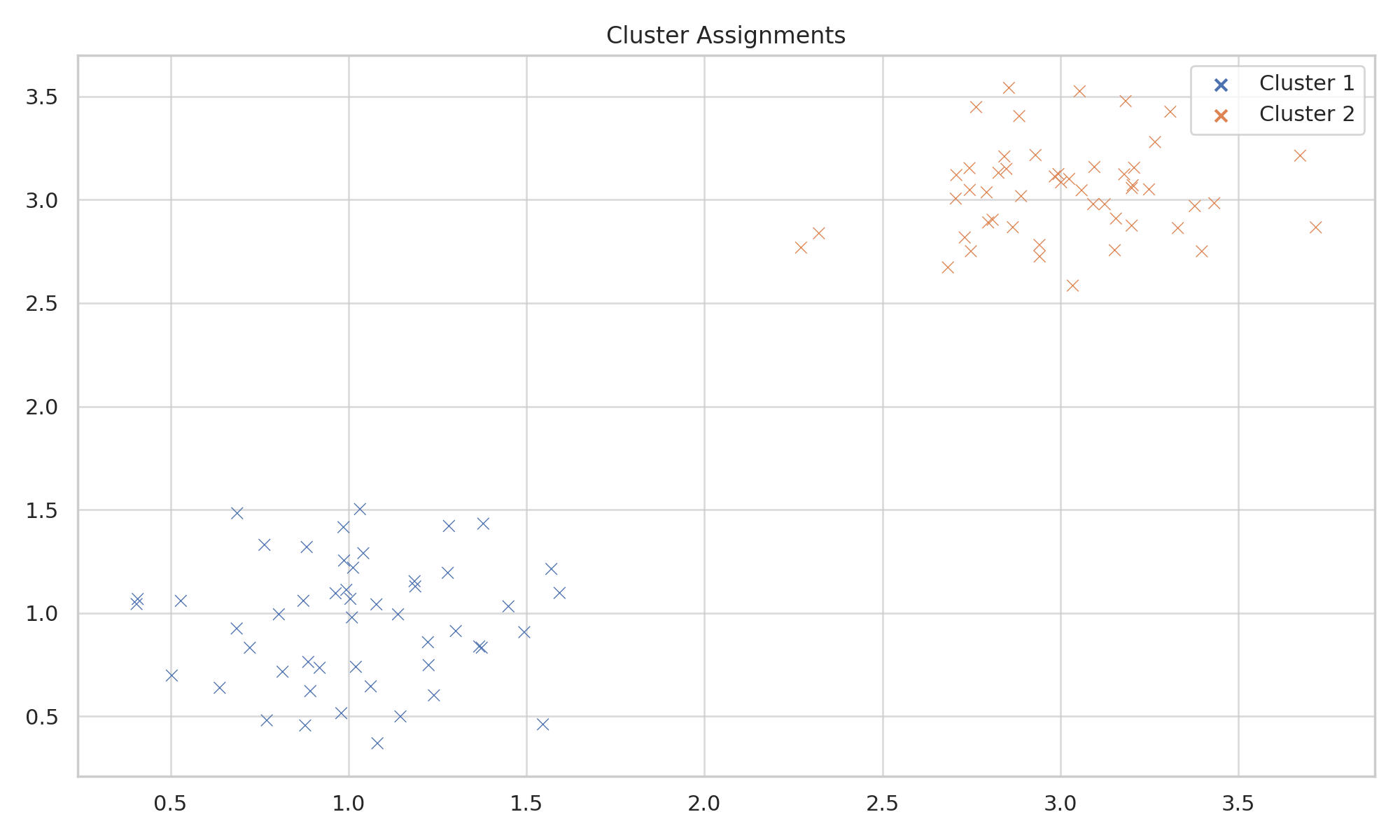}
\caption{Latent Cluster Assignment Visualization}
\label{fig:clusters}
\end{figure}

Figure~\ref{fig:clusters} visualizes the clustering of user embeddings in latent space. Two main clusters are observed, representing typical and anomalous behavior. Each point corresponds to a user’s latent representation, and cluster separation shows the model’s ability to group similar behaviors while isolating suspicious patterns. The use of evidential clustering ensures soft assignments, with uncertain cases often appearing between clusters. Such visualization is valuable for explainability and supports security analysts in understanding model decisions. The clear separation highlights the interpretability benefits of our framework over black-box detectors.

These results confirm that modeling epistemic uncertainty and behavior dynamics significantly improves threat detection fidelity, reliability, and trust.

\section{Conclusion and Future Work}

In this paper, we presented a novel framework for real-time detection of insider threats using behavioral analytics and deep evidential clustering. By combining temporal embeddings of user activity sequences with an uncertainty-aware clustering approach, our model not only achieves high detection accuracy but also significantly reduces false positives—an essential requirement in practical cybersecurity applications. The incorporation of epistemic uncertainty estimation enables better prioritization of alerts and supports adaptive decision-making under ambiguous conditions. Experimental results on two benchmark datasets—CERT and TWOS—demonstrated the superior performance of our approach in terms of accuracy, robustness to concept drift, and interpretability. Our key contributions include the design of a Dirichlet-based clustering head for modeling soft cluster assignments, an anomaly scoring mechanism based on both uncertainty and behavioral drift, and a visualization component for interpretability. These collectively form a robust and deployable system for modern enterprise environments.

For future work, we plan to incorporate active learning mechanisms where high-uncertainty samples are escalated for human labeling to improve detection precision over time. We also aim to expand the behavioral feature space by including additional indicators such as keystroke dynamics, device usage patterns, and cross-platform activity logs, enabling a more holistic profile of user behavior. Furthermore, deploying the framework in real-world SOC environments will allow us to assess long-term adaptability and operational scalability. Lastly, we intend to explore the use of contrastive and self-supervised pretraining techniques to enhance generalization, particularly in low-label or zero-shot scenarios.

Ultimately, our approach bridges the gap between interpretability and performance in insider threat detection, setting a new direction for adaptive and trustworthy AI in cybersecurity.


\end{document}